\documentclass{article}

\usepackage{graphicx}

\begin{document}

%\markboth{Authors' Names}
%{Instructions for Typing Manuscripts (Paper's Title)}

%\catchline{}{}{}{}{}
%
%%%%%%%%%%%%%%%%%%%%%%%%%%%%%%%%%%%%%%%%%%%%%%%%%%%%%%%%%%%%%%%%%%%%

\centerline{\bf Oscillons in Scalar Field Theories: 
Applications in Higher Dimensions
and Inflation}

\vspace{0.5cm}

\centerline{Marcelo Gleiser}

\centerline{\it Department of Physics and Astronomy}
\centerline{\it
Dartmouth College, Hanover, NH 03755, USA\footnote{email: gleiser@dartmouth.edu}}

%\maketitle

%\begin{history}
%\received{Day Month Year}
%\revised{Day Month Year}
%\comby{Managing Editor}
%\end{history}

\begin{abstract}
The basic properties of oscillons -- localized, long-lived, time-dependent
scalar field configurations -- are briefly reviewed, including recent results
demonstrating how their existence depends on the dimensionality of spacetime.
Their role on the dynamics of phase transitions is discussed, and it is
shown that oscillons may greatly accelerate the decay of metastable vacuum
states. This mechanism for vacuum decay -- resonant nucleation -- is then
applied to cosmological inflation. A new inflationary model is proposed which
terminates with fast bubble nucleation.
\end{abstract}

%\keywords{Oscillons; Inflationary Cosmology; Extra Dimensions}

\section{Introduction}

During the past three decades, static, spatially-localized field 
configurations have been of great interest in relativistic field
theories \cite{Rajamaran,Lee}. These nonperturbative solutions of
nonlinear field equations have been shown to exist in a wide variety of models,
with or without nontrivial vacuum topology. Configurations with stability
guaranteed by vacuum topology are called topological solitons. Examples
include kinks, strings, monopoles, and textures \cite{Vilenkin}. When the
configuration's stability comes from conserved charges, they are called
$Q$-balls \cite{Coleman} or nontopological solitons \cite{Lee}.  Both topological
and nontopological solitons have many applications in high-energy particle
physics and  cosmology \cite{Rajamaran,Vilenkin}. Their widespread
appeal  relies on two key properties: first, they are stable, that is, they
maintain their spatial profile through time. (Of course, if the configurations
are allowed to scatter they may or not remain stable and form bound states.)
Second,  their energy density is spatially localized. Thus, they have been used
to model hadrons \cite{Rajamaran,Lee},  as possible signatures of
primordial phase transitions, or to generate the large-scale structure
of the Universe \cite{Vilenkin}.

The purpose of the present work is to present results pertaining to a related
but, in this author's view, equally important class of 
spatially-localized field 
configurations. Their key difference from the solitons mentioned above concerns
their temporal behavior: they are {\it time-dependent}, as opposed to
static, solutions of the equations of motion. What makes them potentially 
interesting for many applications is that,
even though they will eventually radiate their energy to spatial infinity,
the process is quite slow. If their lifetime is longer than
the typical time-scale in the
system they will, for all practical purposes, behave as solitons. 
The remarkable fact about these long-lived,
time-dependent configurations -- called
oscillons -- is that they owe their longevity to the nonlinearities in the
system: no globally-conserved charges or topologically nontrivial boundary
conditions are needed (although they may help extend the oscillon's lifetime).
As such, they can be found in a much broader class of
models.

In the next section oscillons are briefly introduced, followed
by an exploration of their properties in an arbitrary number of spatial
dimensions \cite{doscil}. In section 3, the possible effect of
oscillons on first order phase transitions is discussed \cite{resnuc}. 
Finally, in section 4, they are applied to cosmology, in 
particular to a new two-field
model of inflation that ends with rapid bubble nucleation.

\section{Oscillons in $d$ Dimensions: Basics}

Oscillons were first shown to exist in the context of simple scalar field
theories with symmetric and asymmetric double-well 
potentials \cite{oscil1,oscil2}. 
They were found to be spherically-symmetric,
time-dependent solutions of the equations
of motion
obtained from the $d$-dimensional Lagrangian\cite{doscil},

\begin{equation}
\label{lagrangian1}
L = c_d\int r^{(d-1)}dr \left( \frac{1}{2}\dot\phi^2 - 
\frac{1}{2}\left (\frac{\partial\phi}{\partial r}\right )^2
-V(\phi)\right)~,
\end{equation}
\noindent
where a dot means time derivative.
The $d$-dimensional spatial volume 
element can be written as $d^dx = c_dr^{(d-1)}dr$, where 
$c_d=2\pi^{d/2}/\Gamma(d/2)$ 
is the surface area of a $d$-dimensional sphere of unit radius. An oscillon
is a time-dependent, spatially-localized solution of the
equation of motion $\ddot\phi- \bigtriangledown_d^2 \phi = -V'$, characterized by a nearly
constant energy, $E_{\rm osc}$, and by a persistent nonlinear oscillation
about its core at $r=0$. Think of a rubber sheet that is pinched and
then let go. An oscillon would be a localized oscillation on the
sheet that doesn't get radiated away for a long time. To find
an oscillon, write the scalar field as
\begin{equation}
\label{ansatz}
\phi(t,r) = \left [\phi_c(t) - \phi_v\right ]\exp[-r^2/R^2] + \phi_v~,
\end{equation}
\noindent
where $\phi_c(t)$ is the core value of the field [$\phi(t,r=0)$],
and $\phi_v$ is its asymptotic value at spatial infinity, 
determined by $V(\phi)$. Using the Gaussian profile of eq. \ref{ansatz},
it has been shown that oscillons are present
whenever the energy of the initial configuration $\phi(0,r)$ is larger
than $E_{\rm osc}$ and $R\geq R_{\rm osc}$, where $R_{\rm osc}$ is the
oscillon radius that can be obtained analytically \cite{oscil1}. 
In addition, the potential
must have a portion where $V''<0$ and the field must probe this portion 
at $r\sim 0$. In
fact, the oscillon owes its stability to this ``spinodal instability'' in
the potential \cite{oscil1}.

Writing $A(t)=\phi_c(t)-\phi_v$ and parameterizing the potential as \cite{doscil}
\begin{equation}
\label{potential}
V(\phi) = \sum_{j=1}^{h} \frac{g_j}{j!}\phi^j - V(\phi_v)~,
\end{equation}
\noindent
where the $g_j$'s are constants and the vacuum energy $V(\phi_v)$ is
subtracted from the potential to avoid spurious divergences upon 
spatial integration, one obtains the equation of motion
\begin{equation}
\label{Rconst}
\ddot A = -\frac{d}{R^2}A - \sum_{n=2}^h \left(\frac{2}{n}\right)^{d/2}
\frac{1}{(n-1)!}V^n(\phi_v)A^{n-1}~.
\end{equation}

Expanding the amplitude as $A(t)=A_0(t)+\delta A(t)$ and linearizing
the eom with $\delta A \sim e^{i\omega t}$,
it is found that oscillons may only exist if $\omega^2 <0$. For quartic
potentials with $g_4>0$ (that is, symmetric and asymmetric double wells), 
this conditions implies
\begin{equation}
\label{minimum-radius}
R^2 \geq \frac{d}{\left[\frac{1}{2}\left(\frac{2^{3/2}}{3}\right)^d
\frac{(V''')^2}{V^{IV}}-V''\right]}~.
\end{equation}
That is, oscillons only exist for lumps above a certain critical size. For
$d=3$, $R_{\rm osc} \simeq 2.42$, while for $d=6$, $R_{\rm osc}\simeq 7.5$.
Note that since the denominator of eq. \ref{minimum-radius} must be positive,
there is also an upper critical dimension for oscillons to exist,
\begin{equation}
\label{max-d}
d\leq {\rm Int}\left[\frac{ \ln 2\frac{V''V^{IV}}{(V''')^2}}
{\ln \left(\frac{2^{3/2}}{3}\right)}\right]~.
\end{equation}
For a symmetric double-well, $d_c=6$ \cite{doscil}.

The fact that for these simple models oscillons can only exist below a 
certain number of spatial dimensions raises an interesting possibility.
Let's assume that this model carries the information of the scalar sector of
a more realistic field theory. (Farhi {\it et al.} recently found oscillons
in a $SU(2)$ model \cite{Graham}.] 
Then, if extra dimensions exist and are large \cite{largedim}, 
it is conceivable that oscillons could be
produced in collisions with high enough energy, $E_{\rm col} > E_{\rm osc}$.
Since their lifetime is much longer than the typical time-scales in the
system, they would appear as a late fireball concentrated in a volume
$\sim R_{\rm osc}^d$.
Furthermore, since 
$E_{\rm osc} \sim \frac{1}{2}(\pi/2)^{d/2}d^{d-1}$, 
finding an oscillon would provide information about the
dimensionality of spacetime \cite{doscil}. 
In other words, oscillons could be used as
probes for the existence of higher dimensions.

\section{Resonant Nucleation}

Recently, oscillons have been shown to emerge dynamically when a system is
quenched from a single to a double well potential \cite{res-emerg}. This
means that, under certain conditions, 
they are spontaneously produced during the nonequilibrium 
dynamics of the system. [The same results would be obtained at $T=0$, as long
as the field starts in, say, a Gaussian superposition of momentum modes.]
The system was initially prepared in a thermal state on a single well and then
quenched to a symmetric double well, as is common in Ginzbug-Landau models of
phase transitions \cite{Langer}. The key point is that the system was
prepared so that after the quench the
field was still localized in one of the wells. For example,
the potential could have been changed from $V(\phi)=(\phi+1)^2$ to
$V(\phi)=\frac{1}{4}(\phi^2-1)^2$. The field would then stay trapped around
the $\phi=-1$ minimum.
This is what happens,
for example, during the initial stages of
a first order phase transition, when the field is
initially in a metastable state. More generally, one can write a potential
\begin{equation}
V(\phi)=\frac{m^2}{2}\phi^2-\frac{\alpha}{3}\phi^3+\frac{\lambda}{8}\phi^4~,
\label{V}
\end{equation}
where $\alpha$ is a tunable interaction. Initially, $\alpha=0$ and the
field is in a single well. Then $\alpha$ acquires a positive
nonzero value and the potential becomes a double well. For $\alpha=3/2$
the double well is symmetric; for $\alpha>3/2$ the minimum at
$\phi=0$ is metastable \cite{res-emerg}. 

\begin{figure}
\hspace{2.cm}
\includegraphics[width=245pt,height=220pt]{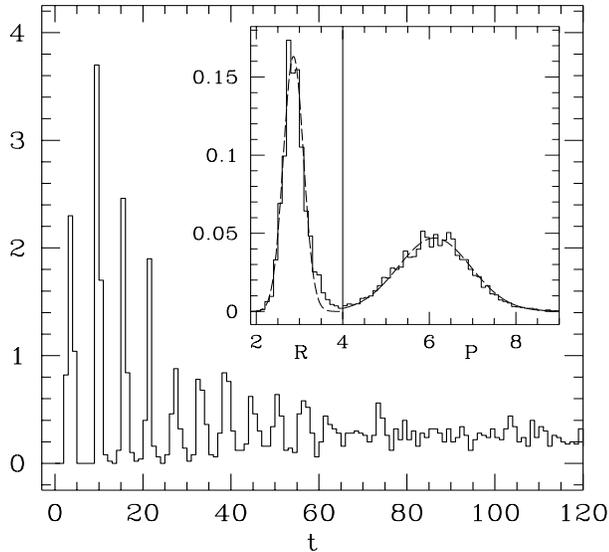}
\caption{ \label{nuc}
The number of oscillons nucleated between $t$ and $t+\delta t$ at $T=0.2$, with
$\delta t=1$. The global emergence is evident early in the simulations. Inset:
the probability distribution of radii and periods of oscillation for the
oscillons nucleated.
}
\end{figure}

The net effect of the quenching is to decrease the effective mass of the
field around the minimum. This key point
may not be apparent at tree level, especially
around $\phi=0$. However, one must remember that
the mass must be corrected due to
thermal and/or quantum fluctuations. One way of incorporating these changes
is to adopt a simple Hartree approximation, shifting the mass from
$m^2$ to $m_H^2=m^2 + \frac{3}{2}\langle\phi^2\rangle$. Before the quench,
the Hartree potential is (upon rescaling the couplings)
$V_H(\phi)=\frac{1}{2}[1+\frac{3}{2}\langle\phi^2\rangle]]\phi^2$.
After the quench,
\begin{equation}
\label{VHartree}
V_H(\phi)=\left [1-m_H^2\right ]\phi +\frac{1}{2}m_H^2\phi^2
-\frac{\alpha}{3}\phi^3+\frac{1}{8}\phi^4~.
\end{equation}
Thus, the quench induces a shift in the minimum from $\phi=0$ to $\phi\simeq
\frac{3}{2}\langle\phi^2\rangle$. This shift, plus the fact that $V_H(\phi)$
acquires a negative derivative at the origin, explains what happens: 
the field's zero mode starts oscillating about the shifted minimum, with
an amplitude controlled both by $\alpha$ and $T$. These oscillations may
induce parametric amplification of a band of $k$-modes which, for a range of
parameters, trigger the emergence of oscillons \cite{res-emerg}. The 
oscillations transfer their energy 
to higher momentum modes via nonlinear scattering and the 
oscillons eventually disappear as the system reaches equipartition.

Fig. 1 shows the number of oscillons nucleated as a 
function of time. Two points are important: first, oscillons emerge 
in synchrony. Second, this initial synchrony is gradually lost
as equipartition is achieved.

\begin{figure}
\hspace{2.cm}
\includegraphics[width=3in,height=3in]{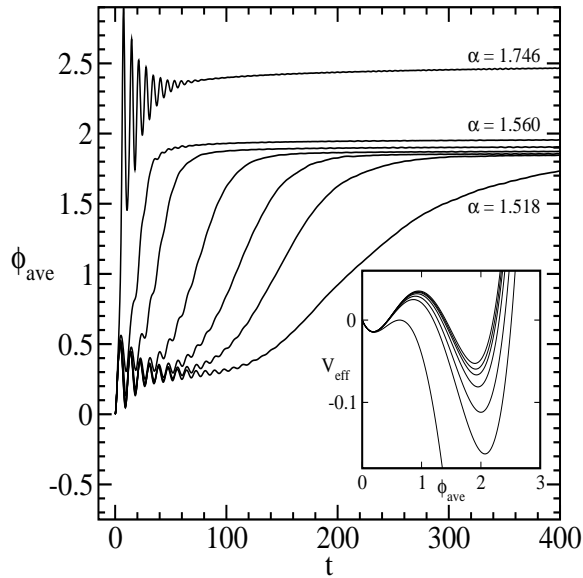}
\caption
{The evolution of the order parameter $\phi_{\rm ave}(t)$ 
at $T=0.22$ for several values of the asymmetry in $d=2$. (From left to right, 
$\alpha= 1.746, 1.56, 1.542, 1.53,
1.524, 1.521, 1.518$. Each curve is an ensemble average over 100 runs. 
The inset shows the effective Hartree potential for the same values of
$\alpha$.}
\label{decay-phi}
\end{figure}

What happens if the potential is asymmetric? In this case, one is
concerned with the way a quenched field will decay from a metastable (or
false vacuum) state to the global minimum. As is well known, if the field is
initially
well-localized about the false vacuum state, it will decay by nucleating
bubbles of a critical size with nucleation rate per unit volume
$\Gamma \simeq T^4\exp[-E(T)/T]$, where $T$ is the temperature and
$E(T)$ is the energy of the critical bubble or bounce 
configuration \cite{Colemanb}. At $T=0$, the quantum decay rate is
$\Gamma\simeq M^4\exp[-S_b]$, where $M$ is the typical mass scale and $S_b$ is
the $d+1$-dimensional Euclidean bounce action. 

As in the case of a symmetric double well, the fast quench will generate a
distribution of oscillons for a certain range of temperatures. The efficiency
of this production mechanism is sensitive to the quenching time scale,
$\tau_{\rm quench}$. If $\tau_{\rm quench} \gg \tau_0$, where $\tau_0$ is the
equilibration time scale of the field's zero-mode, oscillons will be produced.
In Fig. 2 we show the variation of the average value of $\phi$ after the quench.
It can be seen that close to degeneracy ($\alpha \sim 3/2$) the field spends a 
long
time oscillating about the metastable minimum before decaying to the ground
state.

The key point is that a fast enough quench will greatly affect the decay 
rate \cite{resnuc}. 
As shown in Ref. \cite{resnuc}, a fast quench will promote
a much faster decay: for a range of parameters, the decay rate goes from an
exponential to a power law suppression,
\begin{equation}
\label{resnucrate}
\Gamma \simeq T^4[E(T)/T]^{-B}~,
\end{equation}
where $B$ can be obtained numerically. For $d=2$, it was shown that 
$B\simeq 2.9\pm 0.46$
for the range of temperatures where oscillons are copiously 
produced \cite{resnuc}. In Fig. 3 the decay time-scale is plotted against
the bounce energy for different temperatures. It is clear that a power law
is an excellent approximation.

Work in $d=3$ is currently under way. Preliminary results indicate 
that the power law behavior remains, with $B\sim 2$. 

The main consequence of the above result is that whenever the quenching happens
fast, using an exponentially-suppressed decay rate as dictated by false-vacuum
decay theory is simply incorrect. Since $\Gamma\sim \exp[-E(T)/T]$ (or $\sim
\exp[-S_b]$ at $T=0$) 
is widely used in applications ranging from early universe cosmology
(in particular at GUT scales, where $H$ is fast) to models of quark-hadron
phase transitions in collisions such as in RHIC, one should be careful about
their range of validity. It may very well be that transitions that were
considered slow in the past are actually quite fast. As an application
of what was called ``resonant nucleation'' (RN) in Ref. \cite{resnuc}, I
will briefly investigate its possible effects on inflation.

\begin{figure}
\hspace{2.cm}
\includegraphics[width=3in,height=3in]{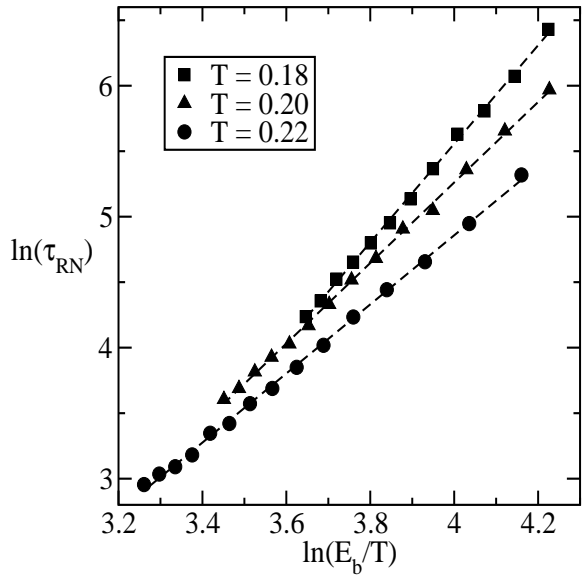}
\caption
{The decay time-scale $\tau_{\rm RN}$ for resonant nucleation as a function 
of critical nucleation 
free-energy barrier $E(T)/T$ at $T=0.18, 0.2$, and $T=0.22$. 
The best fit is a power-law with exponent $B\simeq 2.44$ for $T=0.22$, 
and $B\simeq 3.36$ for $T=0.18$.}
\label{powerlaw}
\end{figure}

\section{Resonant Inflation: Can Old Inflation be Rescued?}

The simple elegance of the original ``Old'' Inflation (OI) scenario proposed by
Guth in 1981 has, since then, inspired many 
variations \cite{Guth,Review}. More than just the elegance of its
formulation, based on a single scalar field decaying from an initial metastable
state to a lower-energy state by bubble nucleation, the original OI model had a
clear connection with particle physics: the inflaton was to be the same scalar
field promoting the symmetry breaking of Grand Unified models, linking
early-Universe cosmology to high-energy particle physics. In fact, it is this
particle physics connection that motivated and motivates the widespread use of
scalar fields in early-Universe physics. 

Unfortunately, Guth's original proposal didn't work. As he himself argued, and
then Linde, and Albrecht and Steinhardt \cite{Linde-new}, the
bubble-nucleation rate could not compete with the exponential expansion rate of
the Universe: the transition would never end. Roughly, while bubble walls
expanded with the speed of light, their centers receded from each other
exponentially fast, making it impossible for the walls to touch, the bubbles to
coalesce, and the transition to complete. Old Inflation gave rise to a universe
with inhomogeneities incompatible with the observed smoothness of the cosmic
microwave background \cite{CMB1}. Guth and Weinberg \cite{GW}, and later
Turner, Weinberg, and Widrow \cite{TWW}, performed a detailed analysis of the
constraints needed to render OI and OI-inspired scenarios viable. They
concluded that a strong (or, equivalently, slow) first order phase transition
based on a single scalar field could not be made to work: the ratio of
decay rate to the expansion rate per unit volume 
[$H^4\simeq (T^2/M_{\rm Pl})^4$], had to be sufficiently small 
\begin{equation}
\label{nobigbub}
\varepsilon \equiv \Gamma/H^4 \leq 10^{-4}~,
\end{equation}
initially,
so that early bubbles didn't produce inhomogeneities during 
nucleosynthesis and on the CMB. (For $\Gamma\simeq T^4\exp[-E(T)/T]$ and 
$T_{\rm GUT}=10^{15}$ GeV, this
implies that $E(T)/T \geq 46.1$ initially).
On the other hand, it had also to grow by the end of inflation
($\varepsilon \rightarrow 9\pi/4$) to guarantee that the transition was 
completed \cite{TWW}. [This implies $E(T)/T \leq 34.9$.]
In other words, successful inflation forced the decay rate to be 
time-dependent: small at the beginning of inflation
and of order unity at the end. 
As further work has shown, this could be achieved by
invoking more fields \cite{HYBINF} and/or a nonminimal gravitational 
coupling \cite{EXTENDEDINF}.

Given what we have learned in the previous section about resonant nucleation,
it is natural to wonder whether such effects can play a role
on inflation. If we write $\varepsilon_{HN}\simeq T^4\exp[-E(T)/T]$ 
to represent the ratio of eq. \ref{nobigbub} using the
homogeneous nucleation rate, and 
$\varepsilon_{RN}\simeq T^4[E(T)/T]^{-B}$
the ratio using the RN rate, equality is attained whenever
\begin{equation}
\label{B}
B = \beta/\ln\beta~,
\end{equation}
where $\beta\equiv E(T)/T$ (or $\equiv S_b$ at $T=0$). 
In Fig. 4 $B$ is shown for
representative values of the nucleation barrier $\beta$. The line denotes
$\varepsilon_{\rm HN}/\varepsilon_{\rm RN} = 1$. The squares denote
the limits imposed by the inflationary constraints of Ref. \cite{TWW}. 
For these values
of $\beta$, unless $B\geq 9$, which is very unlikely, 
resonant nucleation rates are always faster. 
In $d=2$, where $B\simeq 3$,
it is clear that $\varepsilon_{HN}/\varepsilon_{RN}<1$ for all realistic values
of $\beta$, not a surprising result.

\begin{figure}
\hspace{2.cm}
\includegraphics[width=3in,height=3in]{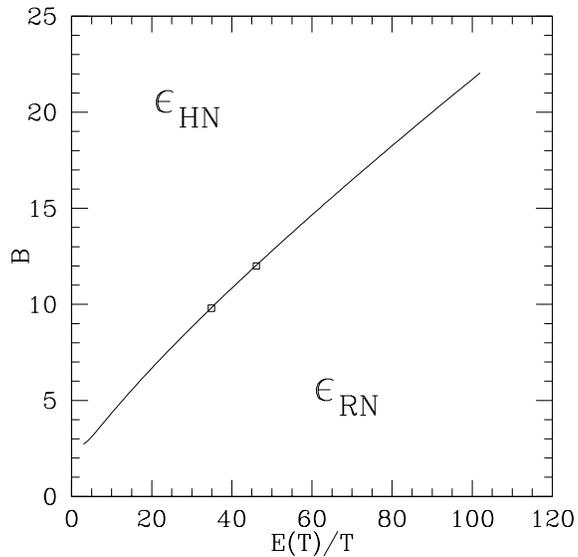}
\caption
{Comparison of homogeneous nucleation (HN) and resonant nucleation (RN) with
power $B$ in an expanding Universe. For a fixed B and nucleation barrier
$\beta=E(T)/T$ (or $S_b$ at $T=0$), the line denotes equality. Values above
the curve imply faster HN, while those below imply faster RN.}
\label{ratio}
\end{figure}

Why is this useful for  inflation? For successful inflation with HN, 
the constraints of Ref. \cite{TWW} limit the nucleation
barrier $\beta$ to be fairly small O($\sim 40$). [See Fig. \ref{ratio}.]
However, calculations of bounce
actions show that $\beta$ usually scales with inverse powers of coupling
constants. These two requirements compete with each
other, making it hard to have small nucleation barriers with small
couplings. Applying the percolation constraint to the RN rate, one
obtains $\beta^B \sim 10^{16}$. For $B=3$, this gives $\beta \sim 10^{16/3}$:
small couplings (or, equivalently, large barriers)
are easier to accommodate with RN, the first reason why
it may be useful for inflation.

The second and most important
reason is that RN makes it much easier to complete the transition.
Clearly, if some mechanism capable of producing the same net
effect as the fast quenching responsible for RN
was present in the early Universe, an initially 
slow first order transition could become fast at some point, 
going from an exponential
to a power law decay. In this way, even a potential with a 
large initial 
barrier would not be an impediment to the successful termination of
inflation. One possible way of implementing RN in cosmology is to invoke
a second field $\psi$ that couples to the nucleating field $\phi$ in a 
way somewhat reminiscent of
hybrid inflation. In that model the inflaton $\phi$
is coupled quadratically to another scalar field $\psi$ which has a 
{\it symmetric} double well potential \cite{HYBINF}:
\begin{equation}
\label{Vhyb}
V(\phi,\psi)=\frac{1}{4\lambda}\left (M^2-\lambda\psi^2\right )^2
+\frac{m^2}{2}\phi^2 + \frac{g^2}{2}\phi^2\psi^2~.
\end{equation}
Inflation is driven by the energy density $V(\phi,0)$
while the inflaton ($\phi)$ is rolling down along the $\psi=0$ 
valley \cite{Review}. 
As $\phi$ reaches a critical value, $\psi$ becomes spinodally unstable
and quickly rolls to one of the minima (or both, but this creates domain
walls, another problem), terminating inflation abruptly. 

The key difference with the mechanism being proposed here
is that bubble nucleation still occurs
at the end of inflation. A possible name is hence
{\it resonant inflation} (RI): it blends OI with the physics
of resonant nucleation.

Modify the potential
for the field $\phi$
that gives rise to RN (e.g. eq. \ref{V}) by coupling another field ($\psi$)
quadratically to it as follows,
\begin{equation}
\label{Vresinf}
V(\phi,\psi)= \frac{1}{2}\left (m^2+g^2\psi^2\right )\phi^2
-\frac{\alpha}{3}\phi^3+\frac{\lambda}{8}\phi^4 + \frac{1}{2}m_{\psi}^2\psi^2 
+|V(\phi_+,0)|~,
\end{equation}
where $\phi_+$ is the value of $\phi$ at the global minimum of $V(\phi,0)$
so that $V(0,\psi)$ provides the net vacuum 
energy responsible for inflation. [Note that here the inflaton is $\psi$.)
Inflation lasts while $\psi$ is rolling down
the $\phi=0$ valley. 
Notice that the mass term for $\phi$, $M_{\phi}^2=m^2+g^2\psi^2$,
decreases as $\psi$ rolls down its potential. While 
$M_{\phi}^2>\alpha^2/2\lambda$, the only minimum in the $\phi$ direction
is at $\phi=0$. However, as $\psi$ decreases, $M_{\phi}^2$ will eventually
drop below $\alpha^2/2\lambda$ and a new minimum will appear at 
$\phi_+=\frac{\alpha}{\lambda}\left [1+(1-2M_{\phi}^2\lambda/\alpha^2)^{1/2}
\right ]$. At $M_{\phi}^2 = \frac{4\alpha^2}{9\lambda}$, 
the two minima are degenerate.
At this point, as $\psi$ continues to approach zero, the minimum at $\phi=0$
becomes metastable. Oscillations in $\phi$, induced by the decrease in its
mass, will induce RN. This will be true as long as the decrease in $M_{\phi}$,
$\dot M_{\phi} \simeq \frac{g^2}{m}\psi\dot\psi $, is fast enough. [It 
was assumed for simplicity
that $g^2\psi^2/m^2 \ll 1$ which is not true for very small $\psi$.]

For RI to work, $\dot M_{\phi}/M_{\phi} < H$ 
during inflation and $\dot M_{\phi}/M_{\phi} > H$ after it. 
During inflation, with a slow-roll approximation, $\psi\dot\psi \simeq
-\frac{m_{\psi}^2}{3H}\psi^2$. We then obtain,
\begin{equation}
\label{slowroll}
\dot M_{\phi}\simeq -g^2\frac{m_{\psi}^2}{3Hm}\psi^2~.
\end{equation}
Also, if $N$ is the number of $e$-folds, 
$\psi_e^2 = \psi_i^2 -\frac{M_{Pl}^2}{2\pi}N$, where $\psi_{i(e)}$ is the
value of the field $\psi$ at the beginning (end) of the inflationary period. 
[For simplicity, it was assumed that during inflation $\frac{1}{2}m_{\psi}^2
\psi^2 > |V(\phi_+,0)|$, that is, inflation is dominated initially by the
vacuum energy of the inflaton field $\psi$.]
Slow-roll ends when $\psi_e^2\leq M_{Pl}^2/12\pi$. Using this result
and eq. \ref{slowroll}, the slow variation of $M_{\phi}$ implies,
$g^2(12\pi)^{1/2}<(m/M_{Pl})^2$. [For $m\sim 10^{16}$GeV, $g < 4
\times 10^{-4}$.] This condition is also consistent with the approximation
$g^2\psi^2/m^2 \ll 1$ for $\psi_i>\psi>\psi_e$, that is, during inflation.

If slow-roll ends when the minimum in $\phi_+$
appears, we obtain 
(this is similar to the critical condition in hybrid inflation \cite{HYBINF}),
\begin{equation}
\label{newmin}
\frac{\alpha_0^2}{2\lambda} = 1 +
\frac{g^2 M_{Pl}^2}{12\pi m^2}~,
\end{equation}
where we defined for convenience $\alpha\equiv m\alpha_0$. The condition
for slow variation of $M_{\phi}$ during inflation forces the second term on
the rhs of eq. \ref{newmin} to be very small.
Thus, if we want to impose that the $\phi_+$ minimum appears close to the
end of inflation, we must have $\alpha_0^2/2\lambda \sim 1$, not a difficult
condition to satisfy.

As inflation ends, $\psi$ will start rolling down fast towards the $\psi=0$
minimum and oscillate around it. Since in this regime, $\dot M_{\phi}/M_{\phi}
\sim (g^2/m^2)\dot\psi\psi$, the rapid motion of $\psi$ will induce the 
time-dependence in $M_{\phi}$ needed to trigger resonant bubble
nucleation. In order
for the transition to end successfully, the percolation constraint
$\varepsilon_{\rm RN} >9/4\pi$, must be satisfied. This
implies,
\begin{equation}
\label{resonantinflation}
\left (S_b\right )^B < \frac{4\pi}{9}\left (\frac{M_{Pl}}{m}\right )^4~.
\end{equation}
If $m\sim 10^{16}$GeV and $B\sim 2$ (as indicated by preliminary results
in $d=3$), RI terminates efficiently if $S_b \leq 10^6$. Since the 
inflationary phase is due to the slow-roll dynamics of the $\psi$ field
and not by the metastable field $\phi$,
the percolation constraint can be satisfied by a wide range of couplings. Also,
since $\phi=0$ only becomes metastable {\it after} 
the end of slow roll, there is
no need to impose the big bubble constraint: once $\psi$ starts rolling
fast at the end of inflation, RN will ensue and 
rapid bubble nucleation and coalescence will quickly reheat the Universe.
Although several details remain to be worked out, this preliminary analysis
indicates that resonant nucleation can be successfully applied to inflationary
cosmology.

\section*{Acknowledgments}

The author thanks Cesar Vasconcelos for the invitation to lecture at the
II International Workshop on Astronomy and Relativistic Astrophysics and for
his good taste in choosing the city of Natal in Brazil for the location.

\end{document}